\RequirePackage{ifpdf}
\ifpdf 
\documentclass[pdftex]{sigma}
\else
\documentclass{sigma}
\fi

\begin{document}

\allowdisplaybreaks

\renewcommand{\PaperNumber}{006}

\FirstPageHeading

\renewcommand{\thefootnote}{$\star$}

\ShortArticleName{Symmetry Extensions in the
Kinetic and Hydrodynamic Plasma Models}

\ArticleName{Symmetry Extensions and Their Physical Reasons \\ in the
Kinetic and Hydrodynamic Plasma Models\footnote{This
paper is a contribution to the Proceedings of the Seventh
International Conference ``Symmetry in Nonlinear Mathematical
Physics'' (June 24--30, 2007, Kyiv, Ukraine). The full collection
is available at
\href{http://www.emis.de/journals/SIGMA/symmetry2007.html}{http://www.emis.de/journals/SIGMA/symmetry2007.html}}}

\Author{Volodymyr B.~TARANOV}

\AuthorNameForHeading{V.B. Taranov}

\Address{Institute for Nuclear Research, 47 Nauky Ave., 03028 Kyiv, Ukraine}
\Email{\href{mailto:wlatar@inet.ua}{wlatar@inet.ua}}
\URLaddress{\url{http://www.geocities.com/vebete/mypage.html}}

\ArticleDates{Received October 31, 2007, in f\/inal form January
14, 2008; Published online January 17, 2008}

\Abstract{Characteristic examples of continuous symmetries in
hydrodynamic plasma theory (partial dif\/ferential equations) and in
kinetic Vlasov--Maxwell models (integro-dif\/ferential equations)
are considered. Possible symmetry extensions conditional and
extended symmetries are discussed. Physical reasons for these
symmetry extensions are clarif\/ied.}


\Classification{35A30; 35Q60; 45K05; 58J70; 22E70}

\section{Introduction}

Symmetry considerations essentially help us to solve nonlinear
problems of plasma physics. Among the recent publications, a
survey of methods which allow us to f\/ind symmetries of
integro-dif\/ferential equations of kinetic plasma theory \cite{[1]} and a
review of applications of symmetry methods to the hydrodynamic
plasma models based on the partial dif\/ferential equations \cite{[2]} can
be cited.

In the present paper, some characteristic examples
of the symmetries  are presented for various plasma theory models: electron
magnetohydrodynamics, collisionless electron plasma oscillations,
drift waves in plasma and multi-component collisionless plasma
containing particles with equal charge to mass ratios.

In Section \ref{sec2}, Lie point symmetries are obtained for the partial
dif\/ferential equations of the electron magnetohydrodynamics by means of the standard Maple~11 program. A~simple invariant solution is
presented, which is in fact a background solution of a perturbation
theory formalism. An additional condition is underlined which can
lead to the symmetry extension.

In Section~\ref{sec3}, continuous symmetries of the integro-dif\/ferential
kinetic equations of the collisionless electron plasma model are
discussed. The symmetries were found in \cite{[3]} by an indirect
algorithm which allows us to obtain symmetries of the kinetic
equations from the symmetries of an inf\/inite set of partial
dif\/ferential equations for the moments of distribution functions.

In Section~\ref{sec4}, some conditional symmetries are presented for the
Hasegawa--Mima hydrodynamic model describing drift waves in a
plasma. Due to these symmetries, invariant solutions are possible
describing the asymptotic structure of the nonlinear waves.

In Section~\ref{sec5}, additional symmetries for integro-dif\/ferential
equations of the kinetic theory of collisionless plasma
containing particles with equal charge to mass ratio are
considered. For example, alpha particles and deuterium ions
participating in a thermonuclear reaction $ {\rm D}^{+} + {\rm T}^{+}
\rightarrow {\rm He}^{++} + n + 17.6$~MeV have close charge to mass
ratios. By simple vector considerations so called extended
symmetry transformations are obtained which allow us to reduce the
number of equations.

Conclusions are made in Section~\ref{sec6}.

\section{Electron magnetohydrodynamics}\label{sec2}

Let us consider the equations of the electron magnetohydrodynamics
(EMHD) \cite{[4]}:
\begin{gather*}
\frac{\partial{\Psi}}{\partial t} =
\nabla\times(\textbf{v}\times\Psi), \qquad \Psi = \textbf{B} -
\Delta {\textbf{B}}, \qquad \textbf{v} = - \nabla\times\textbf{B},
\qquad \nabla\cdot\textbf{B} = 0,
\end{gather*}
where $\Psi$ is the generalized vorticity, $\textbf{B}$ is magnetic
f\/ield strength and $\textbf{v}$ is the hydrodynamic velocity of the
electron plasma component. Nine functions, $\Psi$, $\textbf{v}$,
$\textbf{B}$, of four independent variab\-les~$t$, $x$, $y$, $z$ are present
in this system of partial dif\/ferential equations. Nevertheless,
its symmetry has been obtained by the standard Maple 11 package
program.

The inf\/initesimal operators of Lie point symmetries of the model
are as follows:
\begin{gather*}
X_{1}=\frac{\partial}{\partial t}, \qquad
X_{2}=\frac{\partial}{\partial x}, \qquad
X_{3}=\frac{\partial}{\partial y}, \qquad
X_{4}=\frac{\partial}{\partial z},
\nonumber\\
X_{5}={t}\frac{\partial}{\partial t} -
{\textbf{v}}\frac{\partial}{\partial \textbf{v}} -
{\textbf{B}}\frac{\partial}{\partial \textbf{B}} - {\Psi}
\frac{\partial}{\partial {\Psi}},
\\
X_{6} = \textbf{r}\times\frac{\partial}{\partial \textbf{r}}+
\textbf{v}\times\frac{\partial}{\partial \textbf{v}}+
\textbf{B}\times\frac{\partial}{\partial \textbf{B}}+
{\Psi}\times\frac{\partial}{\partial {\Psi}}.\nonumber
\end{gather*}

The Maple 11 program was tested previously on Korteweg--de Vries,
nonlinear Schr\"odinger, Hasegawa--Mima and other nonlinear plasma
theory models and reproduced well known symmetries. So we can
expect that operators $X_{1}$--$X_{6}$ form a full basis of EMHD
Lie point symmetry algebra.

A little doubt remains, however: maybe some additional symmetries
could be obtained by the direct `manual' use of the standard Lie
algorithm.

Due to time and space homogeneity (symmetries $X_{1}$ to $X_{4}$)
the simplest exact solution exists corresponding to the constant
external magnetic f\/ield which can be directed along the $Oz$ axis
without the loss of generality (because of the rotation symmetry
$X_{6}$). The f\/ield amplitude can be set as $B_{0}=1$ due to the
similarity transform generated by $X_{5}$:
\begin{gather}
\textbf{B}=\textbf{e}_{z}, \qquad \textbf{v}= 0.\label{eq3}
\end{gather}

This solution can be chosen as a background for the perturbation
theory. After the substitution $\textbf{B} \rightarrow
\textbf{e}_{z} + \textbf{B}$, $\Psi \rightarrow \textbf{e}_{z} +
\Psi$ the f\/irst EMHD equation becomes
\begin{gather}
\frac{\partial{\Psi}}{\partial t}
=\frac{\partial{\textbf{v}}}{\partial z}+
\nabla\times(\textbf{v}\times\Psi)\label{eq4}
\end{gather}
but other equations remain unchanged. In this way the system of
equations describing helicon waves in a plasma is usually
obtained.

We can expect that EMHD model is much more symmetric. For example,
nonlinear term in~\eqref{eq4} vanishes under an additional condition:
\begin{gather}
\textbf{v}\times\Psi = \nabla F,\label{eq5}
\end{gather}
where $F$ is arbitrary scalar function. In this way we obtain the
set of linear partial dif\/ferential equations and the nonlinear
algebraic restriction \eqref{eq5}. This property can be a source of rich
additional symmetries.

\section{Collision less electron plasma}\label{sec3}

Let us consider electron collisionless plasma. In this case
the Vlasov--Maxwell integro-dif\/ferential system of equations holds:
\begin{gather}
\frac{\partial f}{\partial t} + {v} \frac{\partial f}{\partial
{x}} - {E} \frac{\partial f}{\partial {v}}
 =0,\label{eq6}
\\
\frac{\partial E}{\partial x} = 1 - \int
_{-\infty}^{\infty} f d {v}, \qquad \frac{\partial
E}{\partial t} = \int_{-\infty}^{\infty} v f d {v},\label{eq7}
\end{gather}
where $f(t,x,v)$ is the distribution function of the electron
component of the plasma, ${E}(t,x)$ is the electric f\/ield strength.

We cannot f\/ind symmetries of this integro-dif\/ferential model by
the traditional Lie method.  This can be done, however, by the
indirect algorithm \cite{[3]}.

First, let us introduce the moments:
\begin{gather}
M_k (t,x)= \int_{-\infty}^{\infty} v^k f d {v},\qquad k =
0, 1, 2, \dots\label{eq8}
\end{gather}
and obtain an inf\/inite set of partial dif\/ferential equations for
them, which is as follows:
\begin{gather*}
\frac{\partial M_k}{\partial t} + \frac{\partial M_{k+1}}{\partial
{x}} + {E} {k} M_{k-1} = 0,\qquad \frac{\partial E}{\partial x} =
1 - M_0, \qquad \frac{\partial E}{\partial t} = M_1.
\end{gather*}

Then, we consider the reduced system involving only f\/irst $N+1$
equations of this set, and we can obtain by the standard Lie procedure
the symmetries of this system:
\begin{gather}
X_{1}=\frac{\partial}{\partial t}, \qquad
X_{2}=\frac{\partial}{\partial x}, \qquad X_{3}={x}
\frac{\partial}{\partial x} + {E} \frac{\partial}{\partial E} +
\sum_{k=1}^{N+1} k M_k \frac{\partial}{\partial M_k},
\nonumber\\
X_{4}=\cos{(t)} \left(\frac{\partial}{\partial x} +
\frac{\partial}{\partial E}\right) - \sin{(t)} \sum_{k=1}^{N+1} k
M_{k-1} \frac{\partial}{\partial M_k},\label{eq10}
\\
X_{5}=\sin{(t)} \left(\frac{\partial}{\partial x} +
\frac{\partial}{\partial E}\right) + \cos{(t)} \sum_{k=1}^{N+1} k
M_{k-1} \frac{\partial}{\partial M_k},\nonumber
\end{gather}
and an additional symmetry
\begin{gather*}
X_{FGN} = F(t) \frac{\partial}{\partial M_N} + \left( G(t) -
{x}\frac{\partial F(t)}{\partial t} \right) \frac{\partial}{\partial
M_{N+1}},
\end{gather*}
$F(t)$ and $G(t)$ being arbitrary functions of $t$.

The standard Maple~11 program can help us to check these results
for not very large $N$ (depending on computer capacity).

Only two highest moments $M_N$ and $M_{N+1}$ are present in
$X_{FGN}$. So only moments with numbers greater than any
previously chosen are involved in the transform generated by
$X_{FGN}$. In the limit $N \rightarrow \infty$ the symmetry group
is generated by $X_{1}$--$X_{5}$.

In general, restoring kinetic symmetries from the symmetries of
the corresponding set of the moments equations is a hard problem
in pure mathematics. Nevertheless, for the relatively simple
operators $X_{1}$--$X_{5}$ it was shown in \cite{[3]} that in the case
considered they correspond to the following inf\/initesimal
operators of the kinetic theory:
\begin{gather}
X_{1}=\frac{\partial}{\partial t}, \qquad
X_{2}=\frac{\partial}{\partial x}, \qquad X_{3}={x}
\frac{\partial}{\partial x} + {v} \frac{\partial}{\partial v} +
{E} \frac{\partial}{\partial E} - {f} \frac{\partial}{\partial f},\label{eq12}
\\
X_{4}=\cos{(t)} \left(\frac{\partial}{\partial x} +
\frac{\partial}{\partial E}\right) - \sin{(t)} \frac{\partial}{\partial
v}, \qquad X_{5}=\sin{(t)} \left(\frac{\partial}{\partial x} +
\frac{\partial}{\partial E}\right) + \cos{(t)}  \frac{\partial}{\partial
v}.\label{eq13}
\end{gather}

Strictly speaking, from the def\/inition \eqref{eq8} it can be readily seen
that the transformations generated by the operators \eqref{eq12}, \eqref{eq13} in
the space of variables $t$, $x$, $v$ and $f$ lead to the transformations
generated by the inf\/initesimals \eqref{eq10} in the space of variables $t$,
$x$ and $M_k$, $k = 0, 1, 2, \dots$.

The symmetries of the same electron collisionless plasma kinetic
model \eqref{eq6}, \eqref{eq7} were found in \cite{[5]} by the direct algorithm, i.e.\
without using of an inf\/inite set of moments equations. The same
generators \eqref{eq12}, \eqref{eq13} were obtained.

\section[Hasegawa-Mima model]{Hasegawa--Mima model}\label{sec4}

Let us consider an inhomogeneous plasma slab in the external
homogeneous magnetic f\/ield. Electrons, unlike ions, are
magnetized, smoothing an electrostatic potential $\Phi$ along the
magnetic f\/ield lines. In this case, Hasegawa--Mima model equations
hold \cite{[6]}:
\begin{gather}
\frac{\partial \Psi}{\partial t} + J(\Phi,\Psi) = \frac{\partial
\Phi}{\partial y}, \qquad \Psi = \Phi - \Delta_\bot
 \Phi, \label{eq14}
\end{gather}
where $\Psi$ is the generalized vorticity,
\begin{gather*}
J(F,G)\equiv \frac{\partial F}{\partial x}\frac{\partial
G}{\partial y} - \frac{\partial G}{\partial x} \frac{\partial
F}{\partial y}, \qquad \Delta_\bot \equiv
\frac{\partial^2}{\partial x^2}+\frac{\partial ^2}{\partial y^2}.
\end{gather*}

The simultaneous presence of the $\Phi$ and $\Delta_\bot \Phi$ in
the second equation of \eqref{eq14} leads to the symmetry reduction and,
as a consequence, to the absence of self-similar solutions \cite{[7]}.

Nevertheless, self-similar solutions can exist as a consequence of
conditional symmetries \cite{[8]}. The notion of conditional symmetry is
discussed and large bibliography is presented in a recent paper
\cite{[9]}.

For example, if an additional condition
\begin{gather}
\frac{\partial^2 {\Phi}}{\partial x^2}=0\label{eq16}
\end{gather}
is fulf\/illed, the term $\Delta_\bot \Phi$ is reduced to
$\frac{\partial^2 \Phi}{\partial y^2}$, which leads to the
symmetry extension. In this way we obtain more symmetric, but
still nonlinear, set of equations, which admits the following
similarity transform:
\begin{gather*}
X_{1}={x} \frac{\partial}{\partial x} +{\Phi}
\frac{\partial}{\partial \Phi}+{\Psi}\frac{\partial}{\partial
\Psi},
\end{gather*}
which allows us to try solutions in the form:
\begin{gather*}
  {\Phi}= x F ( t, y ), \qquad    {\Psi}= x G ( t, y ).
\end{gather*}

$G = - 1$ is a trivial solution which is not interesting for
physics. So only the simple nonlinear exact solution with $G \neq
- 1$ important for physics as a nonlinear wave asymptotic, will be
considered below. More complicated solutions can be presented and
discussed in future papers.

For the functions $F(t,y)$ and $G(t,y)$ we obtain the equations
\begin{gather}
\frac{\partial (1/(G+1))}{\partial t}+ \frac{\partial
(F/(G+1))}{\partial y}=0, \qquad G = F - \frac{\partial^2
F}{\partial y^2}.\label{eq19}
\end{gather}

Equations \eqref{eq19} have a particular solution \cite{[8]}:
\begin{gather}
F = \alpha (1+\beta \cos ((\omega+\delta \omega) t+q y)),\label{eq20}
\end{gather}
so that
\begin{gather}
{\Phi}= \alpha x (1+\beta \cos ((\omega+\delta \omega) t+q y)),\label{eq21}
\end{gather}
where $\alpha$ is arbitrary constant amplitude, constant factor
$\beta$ determines the relative weight of the zonal f\/low and the
monochromatic wave, the frequency is $\omega = q /( 1 + q^2 )$ and
the frequency shift is $\delta \omega = - \alpha q^3 / ( 1 + q^2
)$.

Another class of conditionally symmetric invariant solutions
became possible under the following condition:
\begin{gather*}
\frac{\partial^2 {\Phi}}{\partial y^2}=0.
\end{gather*}

In this case the following similarity transform appears as a
conditional symmetry:
\begin{gather*}
X_{2}={t} \frac{\partial}{\partial t}+{y} \frac{\partial}{\partial
y}
\end{gather*}
and another class of invariant solutions is possible
\begin{gather*}
\Phi = F(x) \frac{y}{t} + G(x), \qquad \Psi = ( F(x) - F''(x) )
\frac{y}{t} + G(x) - G''(x).
\end{gather*}

Pure self similar solutions $G(x) = 0$ are absent in this case,
but there exist solutions corresponding to the arbitrary shear
f\/low $F(x) = 0$, $G(x)$ being an arbitrary function of $x$.

\section{Multi-component collisionless plasma}\label{sec5}

Let us consider $N$-component collisionless plasma. In this case
the Vlasov--Maxwell integro-dif\/ferential system of equations holds:
\begin{gather}
\frac{\partial f_\alpha}{\partial t} + \textbf{v} \frac{\partial
f_\alpha}{\partial \textbf{r}}
+\frac{e_\alpha}{m_\alpha}\left(\textbf{E} + \frac{1}{c} [\textbf{v}
\times \textbf{B}] \right) \frac{\partial f_\alpha}{\partial \textbf{v}}
 =0,\label{eq25}
\\
\nabla\times\textbf{E}+ \frac{1}{c} \frac{\partial
\textbf{B}}{\partial t}
 =0,
 \qquad \nabla\cdot\textbf{E}
 =4\pi\rho,
 \qquad
\nabla\times\textbf{B}= \frac{1}{c} \frac{\partial
\textbf{E}}{\partial t}+\frac{4\pi}{c} \textbf{j}, \qquad
\nabla\cdot\textbf{B}
 =0,\label{eq26}
\end{gather}
where ${f_\alpha}(t,\textbf{r},\textbf{v})$ is the distribution
function of the $\alpha-th$ component of the plasma,
$\alpha=1,\dots,N$. Charge and mass of particles of the $\alpha$th
component are denoted by $e_\alpha$ and $m_\alpha$,
respectively. Charge and current densities have the form
\begin{gather}
\rho=\sum_{\alpha=1}^N e_\alpha \int_{-\infty}^{\infty}
f_\alpha d\textbf{v}, \qquad \textbf{j}=\sum_{\alpha=1}^N e_\alpha
\int_{-\infty}^{\infty} \textbf{v}f_\alpha d\textbf{v}.\label{eq27}
\end{gather}

Let  $\alpha = 1$ and $\alpha = 2$ correspond to plasma components
with equal charge to mass ratio of particles, for example, alpha
particles and deuterium ion components, so we assume (and it is a~good approximation:
\begin{gather*}
\frac{e_{1}}{m_{1}}=\frac{e_{2}}{m_{2}}.
\end{gather*}

According to \eqref{eq27}, these components enter the Maxwell equations
only as the sum
\begin{gather}
{e_{1}}{f_{1}}+{e_{2}}{f_{2}},\label{eq29}
\end{gather}
which is in fact the distribution function of the charge density
of the components   $\alpha = 1$ and  $\alpha = 2$. In addition,
according to \eqref{eq25}, the functions ${f_{1}}$, ${f_{2}}$ and
${e_{1}}{f_{1}}+{e_{2}}{f_{2}}$ satisfy the same Vlasov equation.
So the transformation \cite{[8]}
\begin{gather}
f_{1}'={f_{1}}-{e_{2}}F({f_{1}},{f_{2}}),\qquad
f_{2}'={f_{2}}+{e_{1}}F({f_{1}},{f_{2}}),\label{eq30}
\end{gather}
where $F({f_{1}},{f_{2}})$ is an arbitrary function of its
arguments, leaves the Vlasov--Maxwell equations invariant, at least
if we perform in \eqref{eq27} the summation f\/irst and the integration
later.

The transformation \eqref{eq30} does not preserve, in general, the
positiveness of the distribution functions. Moreover, they can
lead to the divergence of the moments, which can be shown, for
example, by choosing $F = {\rm const}$.

Nevertheless, we can choose $ F(f_{1},f_{2})=f_{1}/e_{2} $ to
obtain the simplifying transformation
\begin{gather}
f_{1}'= 0,\qquad e_{2}f_{2}'=e_{1} f_{1} + e_{2} f_{2}.\label{eq31}
\end{gather}

The simplifying transform \eqref{eq31} can also be obtained from the
symmetry deduced from the symmetry of moments' equations,
\begin{gather*}
f_{1}'={f_{1}}\exp(a),\qquad
f_{2}'={f_{2}}+\frac{e_{1}}{e_{2}}(1-\exp(a))f_{1},
\end{gather*}
$a$ is arbitrary constant, in the limit $ a \rightarrow - \infty$.

According to \eqref{eq31}, we can choose the invariant \eqref{eq29} as a new
distribution function and solve the system \eqref{eq25}, \eqref{eq26} for
\begin{gather*}
{e_{1}}{f_{1}}+{e_{2}}{f_{2}}, \quad {f_{3}}, \quad \dots, \quad {f_{N}}
\end{gather*}
and the f\/ields $\textbf{E}$ and $\textbf{B}$, so the number of
equations is reduced by one.

\section{Conclusions}\label{sec6}

Lie point symmetries of electron magnetohydrodynamics are obtained
(Section \ref{sec2}), which include time and space homogeneity, self
similarity and rotations. As a consequence of the symmetries,
simple invariant solution \eqref{eq3} exists which is in fact the
background solution which allows to explore nonlinear waves
evolution by a perturbation formalism. Rich conditional symmetries
are possible in this model due to additional conditions like \eqref{eq5}.

Lie point symmetries of the integro dif\/ferential Vlasov--Maxwell
model for collisionless electron plasma obtained in \cite{[3]} by the
moments' method, i.e.\ from the symmetry of an inf\/inite set of
partial dif\/ferential equations for the moments of the distribution
functions, are presented in Section~\ref{sec3}.

It is shown in Section~\ref{sec4} that hydrodynamic Hasegawa--Mima model for
drift waves in magnetized plasma, which, in general, is not self
similar, has exact asymptotic self similar solutions~\eqref{eq21}. It is
possible due to the conditional symmetry of the model as a
consequence of an additional condition~\eqref{eq16}. The physical reason
for this symmetry extension is that nonlinear waves have a~simpler
shape near their critical points -- nodes, crests etc.

In Section~\ref{sec5}, by simple vector considerations, it is shown that
multi-component collisionless plasma containing components with
equal charge to mass ratio of particles is invariant under the
transformation \eqref{eq30}, if we perform the summation f\/irst and the
integration later in the expressions for charge and current
densities \eqref{eq27}. This transformation, called the extended symmetry,
does not preserve, in general, the global conditions of non
negativity of distribution functions and existence of their
moments. So, in contrast with the conditional symmetries, this
symmetry extension is a consequence not of imposing additional
conditions, but of neglecting some global constraints.
Nevertheless, among the extended symmetries \eqref{eq30} we can f\/ind the
simplifying transform \eqref{eq31}, which allows us to reduce the number of
equations by~1.

The physical reason for the extended symmetry \eqref{eq30} is that
particles with equal charge to mass ratios under the same initial
conditions (coordinate and velocity) in a given electric and
magnetic f\/ields move along the same trajectories. The second
physically important reason is that we consider very fast
evolution of a plasma, faster than any collisions.

\pdfbookmark[1]{References}{ref}
\LastPageEnding

\end{document}